\documentclass[aps, prd, twocolumn, amsmath, floats,floatfix, superscriptaddress, nofootinbib]{revtex4}

\usepackage{epsfig}
\usepackage{amsmath,amssymb,mathrsfs,mathtools}
\usepackage[caption=false]{subfig}
\usepackage{verbatim}
\usepackage{float}
\usepackage{morefloats}
\usepackage[dvipsnames]{xcolor}
\usepackage{booktabs}
\usepackage[normalem]{ulem}
\usepackage{multirow}
\usepackage{makecell}
\usepackage{tabularx}
\usepackage{cancel}
\usepackage{hyperref}
\usepackage{orcidlink}

\usepackage[T1]{fontenc}

\usepackage[capitalise]{cleveref}

\usepackage{comment}

\crefname{section}{Sec.}{Secs.}
\crefname{table}{Tab.}{Tabs.}
\crefname{figure}{Fig.}{Figs.}
\crefname{equation}{Eq.}{Eqs.}
\crefname{appendix}{Appendix\ }{Appendix\ }

\providecommand{\openone}{\leavevmode\hbox{\small1\kern-3.8pt\normalsize1}}

\definecolor{bostonuniversityred}{rgb}{0.8, 0.0, 0.0}

\newcommand{\high}{} 

\usepackage{xspace}
\usepackage{mathrsfs}
\DeclareSymbolFontAlphabet{\mathrsfs}{rsfs}
\newcommand{\scri}{\mathrsfs{I}}
\newcommand{\scripx}{$\scri^+$}
\newcommand{\scrip}{\scripx\xspace}

\usepackage{booktabs}
\usepackage[Q=yes,pverb-linebreak=no]{examplep}
\usepackage{aas_macros}
\usepackage{scrextend}

\allowdisplaybreaks[4]

\begin{document}

\title{\boldmath Charged Scalar Field at Future Null Infinity via Nonlinear Hyperboloidal Evolution\unboldmath}

\author{Jo\~ao~D.~\'Alvares\,\orcidlink{0000-0001-5501-9014}}
\email{joaodinis01@tecnico.ulisboa.pt}
\affiliation{CENTRA, Departamento de F\'{\i}sica do Instituto Superior T\'{e}cnico (IST), Universidade de Lisboa, 1049-001 Lisboa, Portugal}\
\affiliation{Instituto de Telecomunicaç\~oes (IT), Universidade de Aveiro Campus Universitário de Rua Santiago, 3810-193 Aveiro}

\author{Alex Va\~no-Vi\~nuales\,\orcidlink{0000-0002-8589-006X}}
\email{alex.vano@uib.es}
\affiliation{CENTRA, Departamento de F\'{\i}sica do Instituto Superior T\'{e}cnico (IST), Universidade de Lisboa, 1049-001 Lisboa, Portugal}
\affiliation{Departament de Física, Universitat de les Illes Balears, IAC3, Carretera Valldemossa km 7.5, E-07122 Palma, Spain}

\begin{abstract}
Quasinormal modes and power-law late-time decay tails of a charged scalar field {\high on a charged black hole have been considered in the literature, and here we use the hyperboloidal formalism to present an independent comparison to prior non-linear results.}
{\high For the quasinormal modes we show a fit of the spherical fundamental mode for the purely uncharged case and compare it to the charged one. 
We obtain good agreement with prior prescriptions for the oscillation frequency between real and imaginary parts of the scalar field, as well as for the exponent of the power-law decay.} Full non-linear evolutions of hyperboloidal slices in spherical symmetry were used to obtain our results, allowing for the extraction of signals at future null infinity.\footnote{Note: this version of the paper has been revised to take into account the comments in~\cite{Hod:2025miv}. } 
\end{abstract}

\maketitle

\section{Introduction}



Valuable information about compact objects and their induced spacetime deformations can be retrieved by studying the behaviour of the system after dynamical perturbations. In general relativity, such relaxation takes the form of quasi-normal mode (QNM) ringing~\cite{Kokkotas:1999bd,Konoplya:2011qq,berti_quasinormal_2009, Berti:2025hly}, followed by power-law decay tails~\cite{price_nonspherical_1972,gundlach_late-time_1994,DeAmicis:2024eoy}. Both propagate accross the whole spacetime until infinity, where they are unambiguiously characterized. These effects are variations around a stationary state achieved in the long term, which makes perturbation theory a suitable tool to gain insights into them. A very relevant question concerns the regime of validity of the description by the linear theory. Taking as example the ringdown of a coalescence of compact objects, when after the non-linear merger is the perturbative approach an appropriate description? Only through comparisons between the full non-linear case and linear approximations can the faithfullness of the latter be judged. In this work in spherical symmetry we numerically evolve scalar perturbations of the Reissner-Nortstr\"om (RN) spacetime using the coupled general relativity, Maxwell and complex massless Klein-Gordon systems in the full non-perturbative regime reaching null infinity and compare them to prior estimates in the literature. 

The seminal work \cite{price_nonspherical_1972} motivated a still-growing interest in QNMs and power-law decay tails. 
For the case of a real scalar field, the power-law decay comes from the backscattering of outgoing waves of the scalar field at late-times, due to the effective potential created by the presence of a black hole. Being $l$ the mode of the scalar field $\phi$, in the limit of timelike infinity the expectation is that{\high~\cite{price_nonspherical_1972}}
\begin{equation}
    \phi \propto t^{-{\high (2l+3)}}\, .
    \label{eq:realscalardecay}
\end{equation}
Shortly afterwards, Ref.~\cite{bicak_gravitational_1972} {\high analyzed the late-time power-laws of a neutral scalar field on an electrically charged star collapsing into a RN black hole}. The same power-law as \eqref{eq:realscalardecay} was predicted, but with the signal dominated by slower decaying primary waves. 

About two decades later, Ref.~\cite{gundlach_late-time_1994} studied massless scalar field perturbations in fixed Schwarzschild and RN spacetimes, but this time also focusing on future null infinity -- the collection of the end-points of future-directed null geodesics. Not long after, Hod and Piran \footnote{ We mention the authors' names for ease of comparison with their results in Figs.~\ref{fig:CompRNQNM} and~\ref{fig:CompFreqRNQNM}.} published a three-paper investigation on the late-time evolution of charged gravitational collapse and decay of charged scalar hair \cite{hod_late-time_1998,hod_late-time_1998_II,hod_late-time_1998_III}. In Ref.~\cite{hod_late-time_1998} they agreed with Ref.~\cite{bicak_gravitational_1972}, but only for the regime of $qQ \ll 1$, where $q$ is the scalar field's charge parameter and $Q$ is the black hole's charge. The second paper then corrects the {\high oscillatory} power-law {\high tail} for general values of $qQ$, reading 
\begin{equation}
    \phi \propto {\high e^{i\frac{qQ}{r_+}t} } \ t^{-{\high (}2\beta+2{\high )}}\,
    \label{eq:chargedscalardecaytimelike}
\end{equation}
at timelike infinity{\high, with $r_+=M+\sqrt{M^2-Q^2}$ the event horizon of the RN black hole}. The quantity $\beta$ is defined as
\begin{equation}
    \beta = \frac{-1+\sqrt{(2l+1)^2-4(qQ)^2}}{2}\,.
    \label{eq:chargedscalardecaytime}
\end{equation}
At future null infinity, they also predict that
\begin{equation}
    \phi \propto {\high e^{i\frac{qQ}{r_+}t} }  u^{-(\beta + 1 - iqQ)}\,,
    \label{eq:chargedscalardecaynull}
\end{equation}
where $u$ denotes the retarded time {\high, and is related to $t$ as $t=(u+v)/2$ with $v$ the advanced time}.
The difference between \eqref{eq:realscalardecay} and \eqref{eq:chargedscalardecaytimelike} is that as the parameter $qQ$ increases, the electromagnetic coupling between the charged scalar field's real and imaginary parts becomes stronger and the QNMs that we expect to see start to be dominated by electromagnetism. The same goes for the tail decay. 
{\high Note that~\eqref{eq:chargedscalardecaytimelike} does not reduce to~\eqref{eq:realscalardecay} for $qQ\to0$, because this limit does not commute with $t\to\infty$, as understood in~\cite{hod_late-time_1998}.}
{\high Finally, the numerical non-linear results presented in~\cite{hod_late-time_1998_III} confirm predictions from the earlier papers.}

More recently, Ref.~\cite{konoplya_massive_2013} generalized for the massive charged scalar field in the Kerr-Newman background, but their analysis for the massless charged scalar field is inconclusive. Although the RN spacetime, which we consider here, does not include the presence of angular momentum, it can be used as a toy-model of the Kerr or Kerr-Newman black holes, as it includes the possibility of extremal behaviour when its charge~$Q$ equals its mass~$M$.

The late-decay of the power-law tails can be evaluated along a timelike curve asymptoting to future timelike infinity $i^+$ and along future null infinity $\scri^+$. The former is out of reach for traditional numerical simulations, where the system would need to be evolved forever to reach it. Exceptions are implementations that use a compactification in the time coordinate such as Ref.~\cite{Camden:2025xjv}. Still, evolving for a long time in a typical implementation allows to see the decay along a timelike curve. Observing the decay along future null infinity however requires a different construction. The following options allow to reach it numerically: evolution on single or double null slices~\cite{Winicour:2012znc}, Cauchy-characteristic extraction/evolution \cite{Bishop:1996gt,Taylor:2013zia,moxon_spectre_2021}, Cauchy-characteristic matching \cite{bishop_cauchy-characteristic_1997,ma_fully_2023,Ma:2024hzq} and hyperboloidal evolution \cite{PhysRevLett.10.66, friedrich1983, Hubner:1998hn,Frauendiener:1997zc}. The approach taken in this work is free evolution of the Einstein equations using the BSSN/Z4 formalism \cite{NOK,PhysRevD.52.5428,Baumgarte:1998te,bona_general-covariant_2003} via conformal compactification~\cite{PhysRevLett.10.66} on hyperboloidal slices~\cite{zenginoglu_hyperboloidal_2008}. Previous work on this research line is given by the series~\cite{Vano-Vinuales:2014koa,vano-vinuales_free_2015,Vano-Vinuales:2017qij,Vano-Vinuales:2023yzs,Vano-Vinuales:2023pum,Vano-Vinuales:2024tat} where the Einstein and Klein-Gordon equations were evolved on hyperboloidal slices in spherical symmetry. Recently, Ref.~\cite{alvares_free_2025} extended this to include the Maxwell equations, showing the capability of the formalism to simulate electrically charged spacetimes.

In this paper we present evolutions of a charged scalar field with a charged black hole at the center of the spacetime, going from the purely neutral case ($Q=0$) to the extremal one ($M=Q$) \cite{aretakis_stability_2011}. Since we will be working in spherical symmetry, we will only consider the modes with $l=0$. Given the efficiency of extracting radiation at future null infinity when using hyperboloidal slices, it is easy to check late-time behavior for several values of $qQ$. This will be important to understand the transition of the scalar field's late-time properties from gravitationally-led effects to electromagnetically-led effects.\par

The paper is organized as follows: the Einstein-Maxwell-Klein-Gordon system as implemented in \cite{alvares_free_2025} is briefly described in section~\ref{sec:methods}, followed up by the initial data for the simulations in section~\ref{sec:indata}. The QNM analysis is done in section~\ref{sec:qnm} and the tail regime is discussed in section~\ref{sec:latetime} and~\ref{sec:tail}. For convergence tests of the implementation we redirect the reader to Ref.~\cite{alvares_free_2025}. The conclusions are presented in section~\ref{sec:conclusions}. 

\section{System Setup}
\label{sec:methods}
A detailed presentation of the setup is given in Ref.~\cite{alvares_free_2025}. Here we provide a brief summary.
The system to be considered is described by the following action \cite{lopez_charged_2023, jaramillo_full_2024},
\begin{equation}
    S = \int \left(R - 8\pi\left[(\mathcal{D}_\mu \phi)^*(\mathcal{D}^\mu \phi) \right] - F^{\mu\nu}F_{\mu\nu}\right)\,\sqrt{-g}\,d^4x\;,
    \label{eq:actionfull}
\end{equation}
where $R$ is the Ricci scalar, $\mathcal{D}_\mu = \nabla_\mu + iq A_\mu$ is the gauge invariant covariant derivative, $F_{\mu\nu}$ is the Faraday/electromagnetic tensor, and $\phi$ is the charged complex scalar field, as mentioned before. The equations of motion that rule our system are the Einstein equations, the massless Klein-Gordon equation and Maxwell's equations, given by
\begin{equation}
    \begin{aligned}
        G_{\mu\nu} &= 8 \pi T_{\mu\nu}\,, \\
        \mathcal{D}_\mu \mathcal{D}^\mu \phi &= 0 \,,\\
        \nabla^\mu F_{{\nu}\mu} &= 4 \pi J_\nu\,,
    \end{aligned}
    \label{eq:kgandmax}
\end{equation}
with
\begin{equation}
    J_\mu = \frac{iq}{2}\left[\phi^* \mathcal{D}_\mu \phi - \phi (\mathcal{D}_\mu \phi)^* \right]\,,
    \label{eq:conservedcurrent}
\end{equation}
where the asterisk means the complex conjugate. The Einstein field equations will be evolved using the BSSN/conformal Z4 formalism~\cite{NOK,PhysRevD.52.5428,Baumgarte:1998te, bona_general-covariant_2003} following the approach in \cite{vano-vinuales_free_2015,alvares_free_2025}. 
Contrary to \cite{price_nonspherical_1972,bicak_gravitational_1972,gundlach_late-time_1994,hod_late-time_1998,hod_late-time_1998_II,hod_late-time_1998_III,konoplya_massive_2013}, where the backreaction of the stress-energy tensor on the spacetime is neglected and thus the background spacetime is not evolved, here we consider the fully coupled case.
The stress-energy tensor of this system reads
\begin{equation}
    \begin{aligned}
        T_{\mu\nu} &= F_{\mu\alpha}F_{\nu}^{\alpha} - \frac{1}{4}g_{\mu\nu}F_{\alpha\beta}F^{\alpha\beta}\\
        &+\mathcal{D}_{\mu}\phi \mathcal{D}^{\nu}\phi - \frac{1}{2}g_{\mu\nu}\mathcal{D}^{\alpha}\phi \mathcal{D}_{\alpha}\phi\,.
    \end{aligned}
\end{equation}
We will be assuming spherical symmetry, so of the electric and magnetic fields that the Faraday tensor is usually decomposed into, we will discard the magnetic field and evolve the electric one. See Ref.~\cite{alvares_free_2025} for more details of the setup.

\subsection{Hyperboloidal Slices and Conformal Compactification}

To be able to extract radiation at \scrip, we will use hyperboloidal slices, given by the level sets of a new time variable $t$ related to the physical $\tilde{t}$ as
\begin{equation}
    t = \tilde{t} - h(\tilde{r})\,,
    \label{eq:hypchange}
\end{equation}
where $h(\tilde{r})$ is the height function, whose effect is to raise the slices so they reach \scrip. We also introduce a compactification factor, $\bar{\Omega}$, such that
\begin{equation}
    \tilde{r} = \frac{r}{\bar{\Omega}(r)}\,.
    \label{eq:compacfactor}
\end{equation}
Its purpose is to map the position of \scrip to a given finite radius $r_{\scri^+}$. Finally we do a conformal compactification, such that the components of the rescaled metric $\bar{g}_{\mu\nu}$ do not diverge at \scrip,
\begin{equation}
    \bar{g}_{\mu\nu} = \Omega^2 \tilde{g}_{\mu\nu}\,.
    \label{eq:conftransf}
\end{equation}
The conformal factor $\Omega$ is chosen to be
\begin{equation}
    \Omega = -K_{\text{CMC}}\frac{r^2_{\mathscr{I}^+}-r^2}{6r_{\mathscr{I}^+}}\,,
\end{equation}
where $K_{\text{CMC}}$ is a (negative) constant. Initially it coincides with the trace of the physical extrinsic curvature, providing constant-mean-curvature slices.\par

\subsection{Einstein Field Equations Formulation}

From now on, variables with a tilde on top will refer to the physical quantities, while those with a bar or nothing on top will correspond to conformally rescaled ones.
Applying the 3+1 decomposition to the conformally compactified metric, we define the spatial part of the metric $\bar{\gamma}_{ij}$. In the BSSN formalism it is decomposed into a spatial conformal factor $\chi$ and a conformally rescaled spatial metric $\gamma_{ij}= \chi\bar{\gamma}_{ij}$.
The line element in this formulation is usually written as
\begin{equation}
    \begin{aligned}
    ds^2 = -\alpha^2 dt^2 + \chi^{-1}\gamma_{ij}(dx^i + \beta^i dt)(dx^j + \beta^j dt)\,,
    \end{aligned}
    \label{eq:lineelement}
\end{equation} 
where $\alpha$ is the lapse and $\beta^r$ is the shift, the gauge quantities that are evolved besides $\chi$ and $\gamma_{ij}$. The conformal extrinsic curvature tensor $\bar{K}_{ij}$ is divided into its trace and trace-free parts,
\begin{equation}
    A_{ij} = \chi \bar{K}_{ij} - \frac{1}{3}\gamma_{ij}\bar{K}\,,
\end{equation}
where $\bar{K} = \bar{\gamma}^{ij}\bar{K}_{ij}$. Instead of using $\bar{K}$, we will instead define this new variable $K$,
\begin{equation}
    K = \bar{K} - 2 \Theta\, ,
    \label{eq:kdefinition}
\end{equation}
where $\Theta$ is a quantity belonging to the Z4 part of the formulation. Lastly,
\begin{equation}
    \Lambda^i = \Delta \Gamma^i + 2 \gamma^{ij}Z_j\,,
\end{equation}
with $\Delta \Gamma^i = \Gamma^i - \hat{\Gamma}^i$ and $\Gamma^i = \gamma^{jk}\Gamma^i_{jk}$. $\hat{\Gamma}_{jk}^i$ is the connection coming from a time-independent spatial metric $\hat\gamma_{ij}$ of our choosing, which we set to the flat spatial metric. From the Einstein equations, we evolve $\gamma_{ij}$, $\chi$, $\alpha$, $\beta^i$, $A_{ij}$, $\tilde{K}$, $\Lambda^i$ and $\tilde{\Theta}$. The physical $\tilde{K}$ and $\tilde{\Theta}$ are related to their conformal counterparts as,
\begin{equation}
    \tilde{K} = \Omega K - \frac{3 \beta^i \partial_i \Omega}{\alpha}\,,\quad \tilde{\Theta} = \Omega \Theta\,.
    \label{eq:ktransformations}
\end{equation} 
In spherical symmetry only the scalar quantities and the radial components of vectors and tensors are evolved.

\subsection{Charged Scalar Field}
Adapting to the notation mentioned in the previous section, the physical scalar field now appears with a tilde on top, $\tilde{\phi}$. It is useful to define the rescaled scalar field $\bar{\phi}$ from the physical one $\tilde{\phi}$:
\begin{equation}
    \bar{\phi} = \frac{\tilde{\phi}}{\Omega}\,,
    \label{eq:scalarfielconf}
\end{equation}
From this, the definitions for the real and imaginary parts of the scalar field are,
\begin{equation}
    \bar{\phi} = \bar{c}_\phi + i \bar{d}_\phi \;, \quad \bar{c}_\phi, \bar{d}_\phi \in \mathbb{R} \;\;,
\end{equation}
such that the modulus of the scalar field is defined as
\begin{equation}
    |\bar{\phi}| = \sqrt{\bar{c}_\phi^2 + \bar{d}_\phi^2}\,.
    \label{eq:scalarfmodulus}
\end{equation}
The corresponding time derivatives of both parts of the scalar field are also defined as\sout{,}
\begin{equation}
    \bar{c}_\Pi = \partial_t \bar{c}_\phi, \; \bar{d}_\Pi = \partial_t \bar{d}_\phi . 
\end{equation}

\section{Initial Data}
\label{sec:indata}
As indicated in \cite{alvares_free_2025}, the simplest condition to initially satisfy the momentum constraint, having a scalar field perturbation in $\bar{c}_\phi$, is 
\begin{equation}
    \bar{c}_\Pi = \beta^r\left(\bar{c}_\phi^\prime + \bar{c}_\phi\frac{\Omega^\prime}{\Omega}\right)\, ,
    \label{eq:cpiinitial}
\end{equation}
where $\beta^r$ is the radial component of the shift and the prime indicates a spatial derivative ($\phi^\prime = \partial_r \phi$). However, when performing simulations with these initial data, it was hard to reach the tail regime with the expected exponent. Therefore, we followed Ref.~\cite{vano-vinuales_free_2015} and changed the initial data such that we had the freedom to choose a more ingoing or outgoing pulse, but still solving the constraints initially. To do so, we add a new variable $\psi_A$, such that the initial value for $A_{rr}$ changes to,
\begin{equation}
    {A_{rr}}_0\psi^{-6} \rightarrow ({A_{rr}}_0 + \psi_A)\psi^{-6}\,,
    \label{eq:psiA}
\end{equation}
where $\psi$ is the variable that we use to solve the Hamiltonian constraint for and $\psi_A$ the variable that we solve the momentum constraint for. The substitution performed for $\chi$ is
\begin{equation}
    \chi \rightarrow \chi_0 \psi^{-4}\,.
\end{equation}
Additionaly, in Ref.~\cite{vano-vinuales_free_2015} a helper variable is defined,
\begin{equation}
    \breve{\Pi} = \frac{\gamma_{\theta\theta}\sqrt{\gamma_{rr}}}{\alpha \chi^{3/2}}\left(\dot{\tilde{\Phi}}-\beta^r \tilde{\Phi}^\prime \right)\,,
    \label{eq:helpervar}
\end{equation}
where the dot denotes a time derivative ($\dot{\phi}=\partial_t \phi$). Expressing \eqref{eq:helpervar} in our variables yields
\begin{equation}
    \breve{\bar{c}}_\Pi = \frac{\gamma_{\theta\theta}\sqrt{\gamma_{rr}}}{\alpha \chi^{3/2}}\left(\bar{c}_\Pi - \beta^r \left(\bar{c}_\phi^\prime + \bar{c}_\phi \frac{\Omega^\prime}{\Omega}\right)  \right)\,,
\end{equation}
where $\breve{\bar{c}}_\Pi$ is our new helper variable. As initial data for it, we choose,
\begin{equation*}
    {\breve{\bar{c}}_\Pi}_0 = \sigma \frac{\gamma_{\theta\theta0} \sqrt{\gamma_{rr0}}}{\alpha_0 \chi_0^{3/2}}\beta^r \left(\bar{c}_{\phi0}^\prime + \bar{c}_{\phi0} \frac{\Omega^\prime}{\Omega}\right)\,,
\end{equation*}
with $\sigma$ a real number usually taken to be $\pm 1$. Reverting back the changes from the previous equations, we get that the initial data for $\bar{c}_{\Pi 0}$ is
\begin{equation}
    \bar{c}_{\Pi0} = \beta^r \left(\bar{c}_\phi^\prime + \bar{c}_\phi \frac{\Omega^\prime}{\Omega}\right)\left(1+\frac{\sigma}{\psi^6}\right)\,.
    \label{eq:cpi2}
\end{equation}
The difference between the above and \eqref{eq:cpiinitial} is the last factor involving $\psi$ and $\sigma$.
The sign of $\sigma$ will imply a mostly ingoing or outgoing, depending on whether it takes the value of -1 or 1. In this paper, we will use $\sigma = 0$ for a longer QNM ringing (which is equivalent to \eqref{eq:cpiinitial}), and $\sigma = -1$ for the tail regime analysis.
The initial data for the rest of the variables is available in \cite{alvares_free_2025}.\par

\begin{figure}[h]
    \centering
    \includegraphics[width=0.45\textwidth]{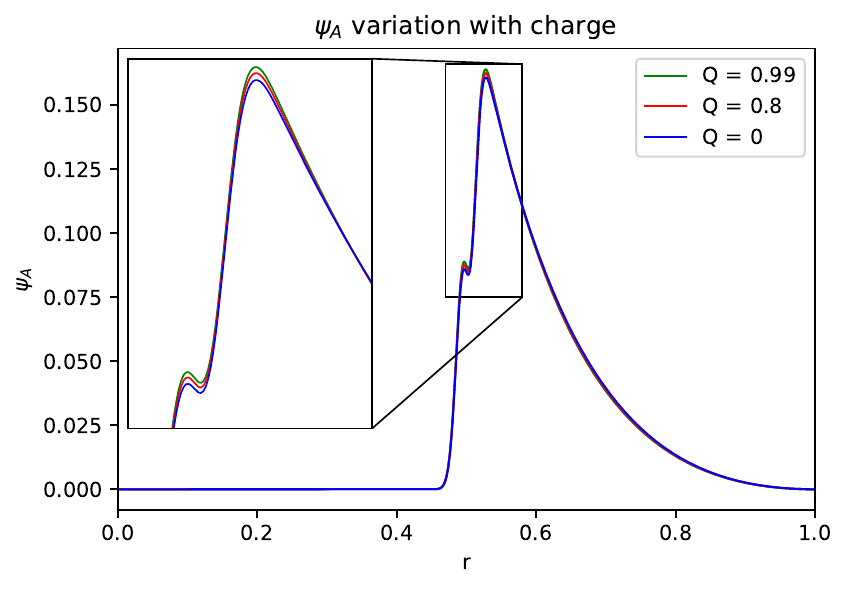}
        \caption{Numerical solution for $\psi_A$ solving \eqref{eq:psiaev} for several values of the black hole's charge. We show a zoom-in on the region where it is easier to see the effect of the charge on $\psi_A$. $M$ is set to 1 and $K_{\text{CMC}}$ to -1.}
    \label{fig:PsiAChange}
\end{figure}

Going back to \eqref{eq:psiA}, to solve the momentum constraint $\psi_A$ has to fulfill the following differential equation 
\begin{equation}
    \begin{aligned}
    \psi_A^\prime &= -\frac{8 \pi \bar{\Omega}^3 {\breve{\bar{c}}}_0 (\bar{c}_\phi^\prime \Omega + \bar{c}_\phi \Omega^\prime)}{\Omega^3}\\
    & - \psi_A\Bigg(\frac{3}{r}\Bigg[\Bigg( 1 - \frac{2M\bar{\Omega}}{r} + \frac{Q^2 \bar{\Omega}^2}{r^2} \Bigg) + \\
    &+ \Bigg(\frac{K_{\text{CMC}}r}{3\bar{\Omega}} + \frac{C_{\text{CMC}}\bar{\Omega}^3}{r^2} \Bigg)^2\Bigg]^{1/2} + \frac{\Omega^\prime}{\Omega}\Bigg)\,,
    \end{aligned}
    \label{eq:psiaev}
\end{equation}
with $C_{\text{CMC}}$ being a constant. Note that this equation is independent of $\psi$. We show how the solution looks for several values of the charge in Fig.~\ref{fig:PsiAChange}, keeping $M=1$ and $K_{\text{CMC}} = -1$. $C_{\text{CMC}}$ is chosen to be the critical value corresponding to the values of $M$, $K_{\text{CMC}}$ and $Q$ that gives us trumpet initial data -- see subsection 3.3.2 in~\cite{vano-vinuales_free_2015} for more details. Furthermore, we chose the scalar field perturbation on $\bar{c}_\phi$ to be Gaussian-like, centered at $\mu = 0.5$, with $\sigma = 0.1$ and the amplitude $A$ to be 0.001:
\begin{equation}
    \bar{c}_{\phi 0} = r^2 A \exp\left(-\frac{(r^2-\mu^2)^2}{\sigma^4}\right) \, .
    \label{eq:cphipert}
\end{equation}

\begin{figure}[h]
    \centering
    \includegraphics[width=0.45\textwidth]{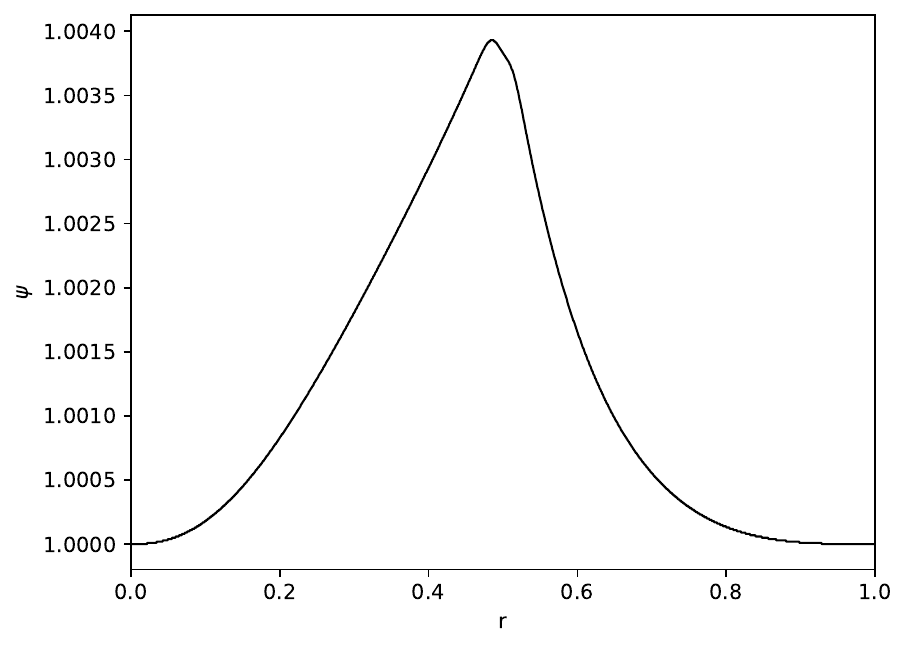}
        \caption{Integrating \eqref{eq:psifrompsia} to get the numerical solution for $\psi$, assuming $Q=0$. For other values of charge, the result is similar.}
    \label{fig:PsiAandPsi}
\end{figure}
The equation for $\psi$, when solving the Hamiltonian constraint only for a perturbation in $\bar{c}_\Pi$ is given by
\begin{equation}
    \begin{aligned}
        \psi^{\prime\prime} &= -\frac{3 C_{\text{CMC}}^2 \bar{\Omega}^4}{4r^6 \psi^7} - \frac{K_{\text{CMC}}^2 \psi}{12 \bar{\Omega}^2}+ \frac{Q^2 \bar{\Omega}^2 \psi}{4r^4}\\ 
        &+\frac{3 C_{\text{CMC}}^2 \bar{\Omega}^4 \psi}{4 r^6} +\frac{K_{\text{CMC}}^2 \psi^5}{12 \bar{\Omega}^2} - \frac{\pi \bar{\Omega}^4 {\breve{\bar{c}}}_0^2}{\psi^7 \Omega^4}\\
        &+\frac{C_{\text{CMC}} \bar{\Omega} \psi_A \Omega}{4r^3 \psi^7} - \frac{3 \psi_A^2 \Omega^2}{16\Omega^2 \psi^7}-\pi \psi \Omega^2 (\bar{c}_\phi^\prime)^2 - \frac{\psi^\prime}{r}\\
        &-\frac{\psi^\prime}{3r^3 \bar{\Omega}}\bigg(K_{\text{CMC}}^2 r^6 + 9 r^4 \bar{\Omega}^2 + 9Q^2 r^2 \bar{\Omega}^4\\
        &+6(C_{\text{CMC}} K_{\text{CMC}} - 3M)r^3 \bar{\Omega}^3 + 9 C_{\text{CMC}}^2 \bar{\Omega}^6\bigg)^{1/2}\\
        &-2\pi \bar{c}_\phi \psi \Omega \bar{c}_\phi^\prime \Omega^\prime - \pi \bar{c}_\phi^2 \psi {\Omega^\prime}^2\,.
        \label{eq:psifrompsia}
    \end{aligned}
\end{equation}
The result from integrating \eqref{eq:psifrompsia}, including the previously obtained numerical solution for $\psi_A$, is shown in Fig.~\ref{fig:PsiAandPsi}, assuming $Q=0$. For other values of the charge, the end result is visually indistinguishable.

\section{Quasinormal Modes}
\label{sec:qnm}

\begin{figure}[h]
    \centering
    \includegraphics[width=0.48\textwidth]{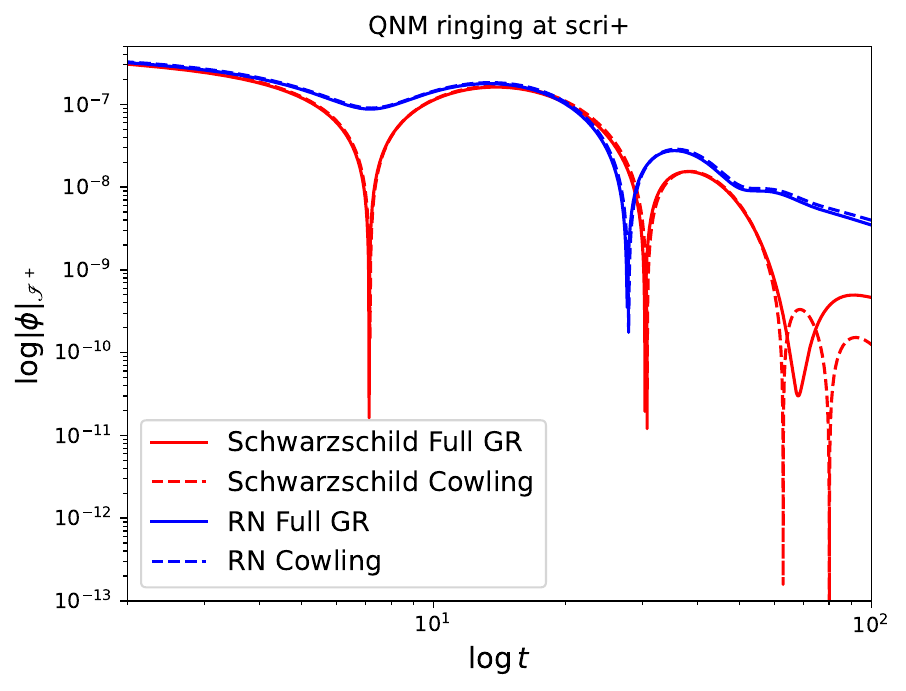}
        \caption{Comparison of quasinormal mode ringing of a scalar field perturbation of Schwarzschild and RN spacetimes (called \textit{Full GR} in the plot). We also show the Cowling approximations, i.e., not evolving the background (called \textit{Cowling} in the plot). We plot this and the following time-dependent graphs in terms of the logarithm of the unrescaled code time $t$ instead of $\log(t/M)$, as the total mass of the system changes depending on the contribution of the scalar field that falls into the black hole.}
    \label{fig:RNQNM_several}
\end{figure}

It is easiest to observe the QNM ringing when the charges of the black hole and the scalar field are set to 0, which corresponds to the usual Klein-Gordon field in Schwarzschild spacetime. Recall that in this section we use \eqref{eq:cpiinitial}, as done in \cite{alvares_free_2025,vano-vinuales_free_2015}. We have further set the following parameters for the real part of the scalar field perturbation \eqref{eq:cphipert}:
$A = 0.0005$, $\mu = 0.4$ and $\sigma = 0.05$. The QNM ringing is shown in Fig.~\ref{fig:RNQNM_several}, together with the Cowling approximation \cite{cowling_non-radial_1941, maniopoulou_traditional_2004}, which corresponds to fixing the background and not evolve the Einstein equations. We also show the same setup for a charged black hole with $Q=0.5$ and setting the scalar field's charge to $q=1.0$.

\begin{figure}[h]
    \centering
    \includegraphics[width=0.48\textwidth]{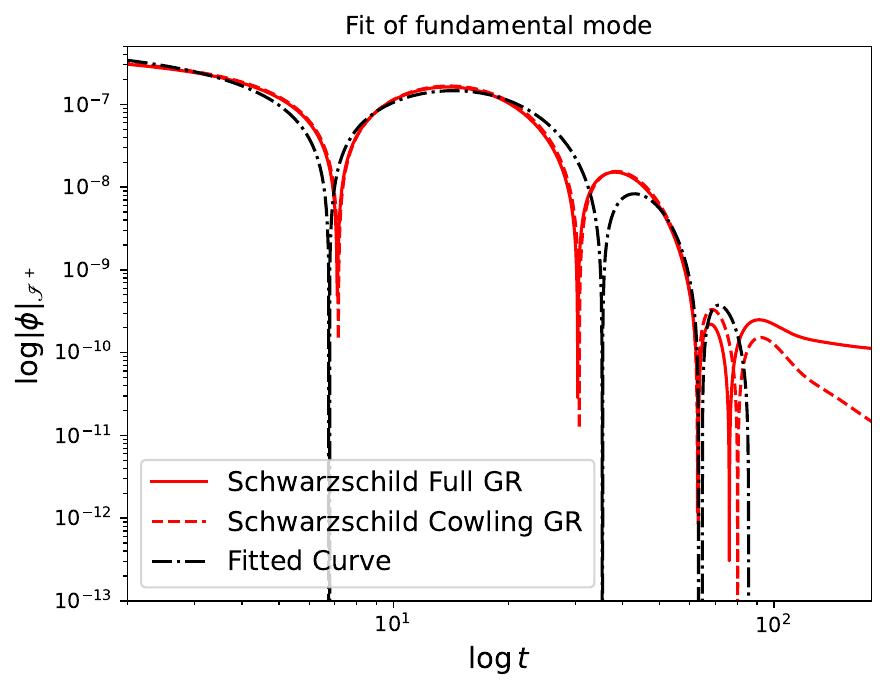}
        \caption{Fitting the fundamental mode to the QNM ringing of the real scalar field, in a Schwarzschild spacetime.}
    \label{fig:schqnm_fit}
\end{figure}

Only in the Schwarzschild case were we able to fit the fundamental spherical mode, because the charged case does not show enough oscillations to be able to estimate their period. Following Refs.~\cite{peterson_spherical_2024,berti_quasinormal_2009}, we assume the shape of the mode to be described by a damped oscillating exponential $e^{-i \omega t}$, with $\omega \approx 0.11 - 0.10 i$ and show the fitting to the data from the simulation in Fig.~\ref{fig:schqnm_fit}. This coincides with the result obtained in~\cite{Peterson:2025csg}.

For sufficiently strong electromagnetic coupling between both objects ($qQ > 0.001$), the amount of oscillations coming from the QNM ringing is indeed too small to do a reliable fit. 
We suspect that the cause of the almost absent ringing is that as soon as the imaginary part picks up on the initial real perturbation, both real and imaginary parts start to oscillate at a very precise frequency (this will be discussed in detail in section~\ref{sec:latetime}), not allowing the overall amplitude of the scalar field to have the same behavior as it would in the purely real case.

\section{Late-time behavior}
\label{sec:latetime}
\begin{figure*}[t]
    \centering
    \subfloat{\includegraphics[width=0.45\textwidth]{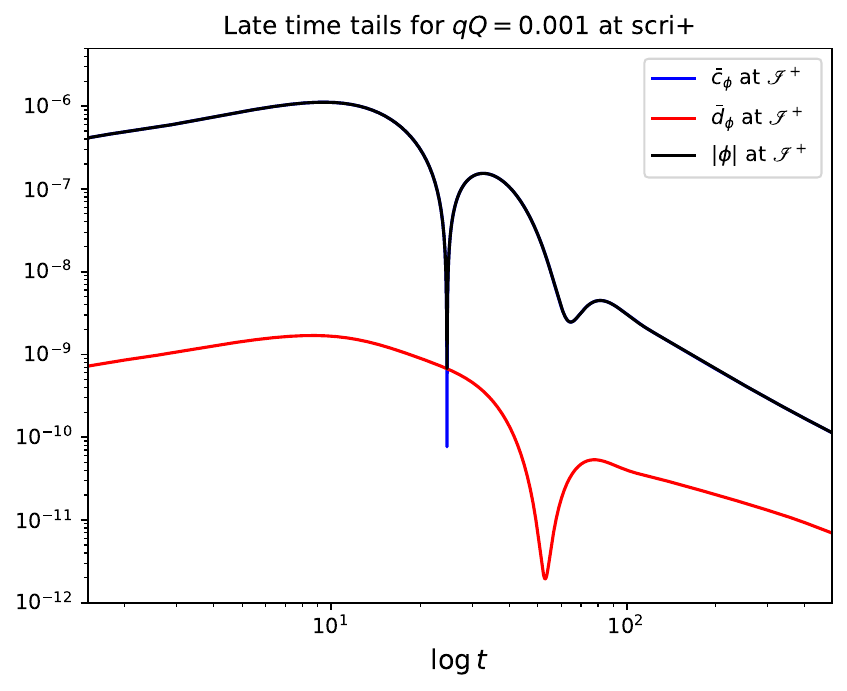}}
    \subfloat{\includegraphics[width=0.45\textwidth]{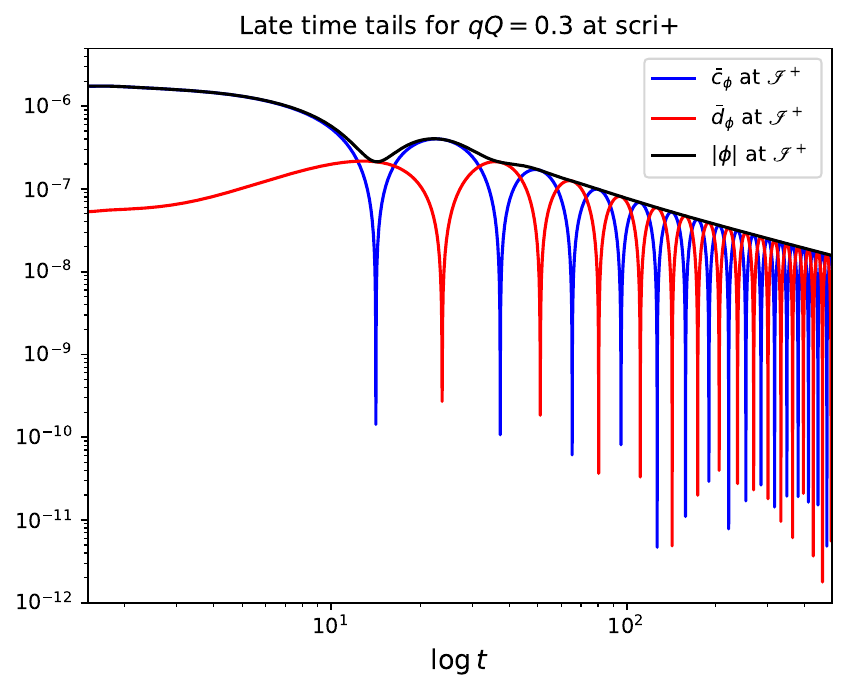}}\\[-2mm]
    \subfloat{\includegraphics[width=0.45\textwidth]{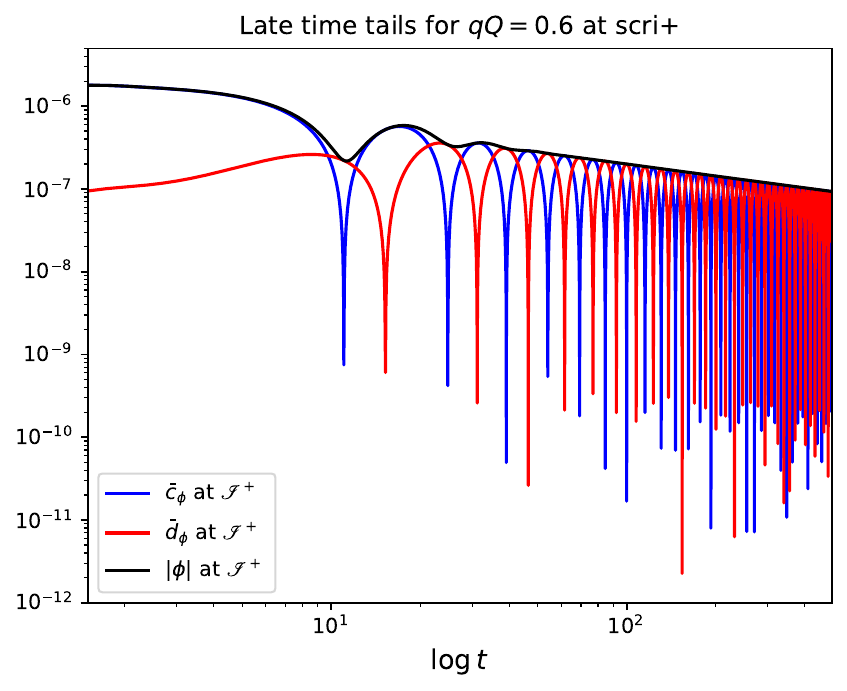}}
    \subfloat{\includegraphics[width=0.45\textwidth]{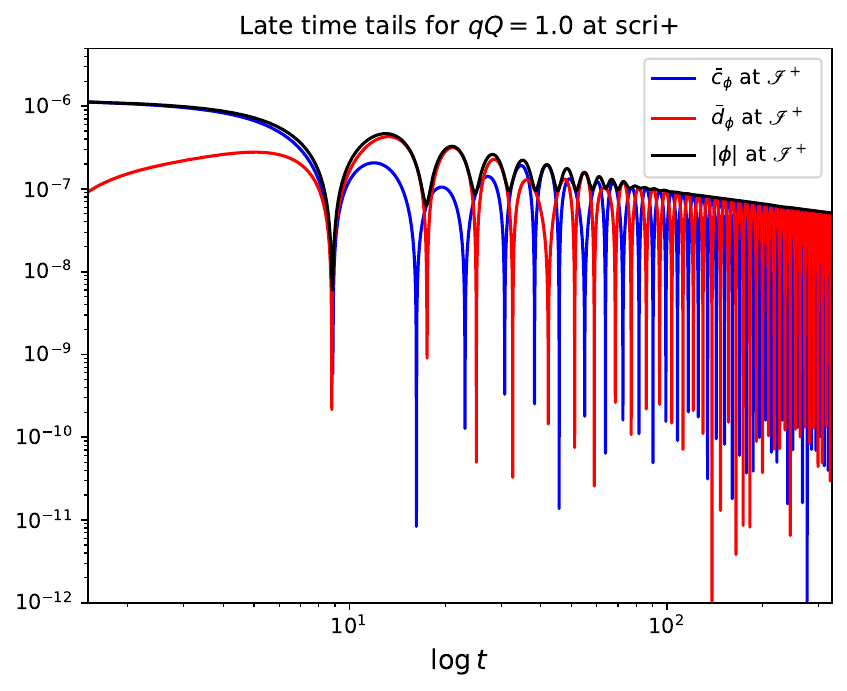}}
    \caption{Late-time tails at \scrip for a charged scalar field in a charged background. The scalar field's charge is set to 1 and the black hole's charge is different in each plot (from top left to bottom right: 0.0001, 0.3, 0.6 and 1.0). The black line in all graphs corresponds to the amplitude of the scalar field \eqref{eq:scalarfmodulus}. The mass of the black hole is set to 1, making the last graph correspond to an extreme RN black hole.}
    \label{fig:RNtail}
\end{figure*}

To check the late-time behavior of the scalar field's perturbation of a RN black hole (after the QNM ringing), we now use the initial data for $\bar{c}_\Pi$ in \eqref{eq:cpi2}. The scalar field's charge is set to 1.0. Regarding the charge of the black hole $Q$, we have ran several simulations, each with a different value of $Q$, to see the effect of the parameter $qQ$ \cite{bicak_gravitational_1972, konoplya_massive_2013,hod_late-time_1998} on the exponent of the power-law regarding the decay. The results are shown in Fig.~\ref{fig:RNtail}. The relevant quantity to look at is the amplitude of the scalar field \eqref{eq:scalarfmodulus}.
For low values of $qQ$ (i.e., the top left graph), both the real and imaginary parts enter a power-law decay, mostly dominated by the real part. However, as we increase $qQ$, the real and imaginary parts continue to oscillate one around the other, presumably due to the increased electric coupling between each other. Even during their oscillation, their amplitude decreases with a fixed power-law tail. Note also that the frequency of oscillation during the tail regime increases with $qQ${\high, see Fig.~\ref{fig:CompFreqRNQNM} and corresponding explanation for further details}.

\section{Tail Regime}
\label{sec:tail}
\begin{figure}[h]
    \centering
    \includegraphics[width=0.45\textwidth]{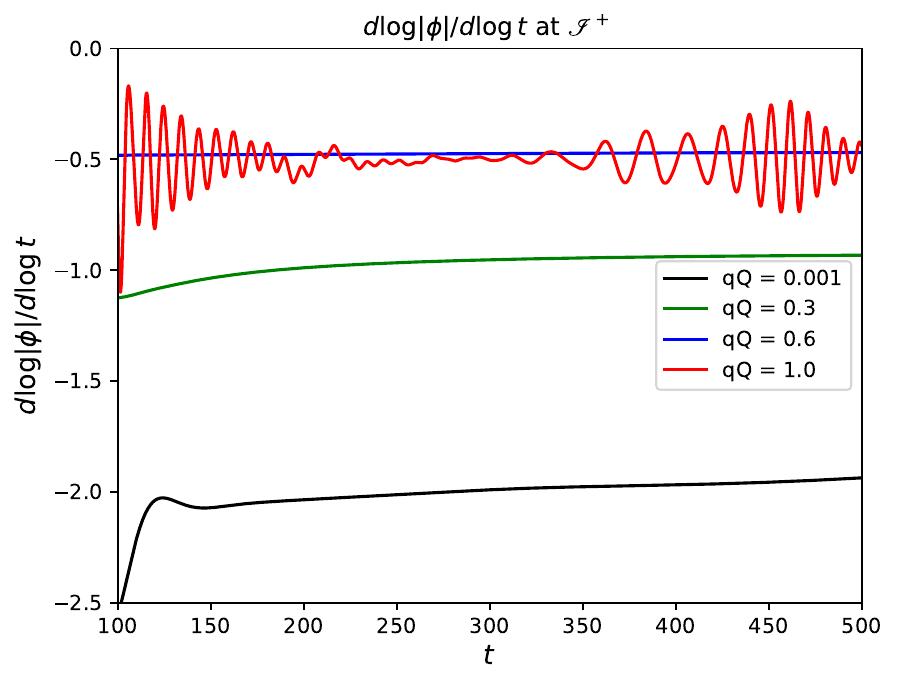}
    \includegraphics[width=0.45\textwidth]{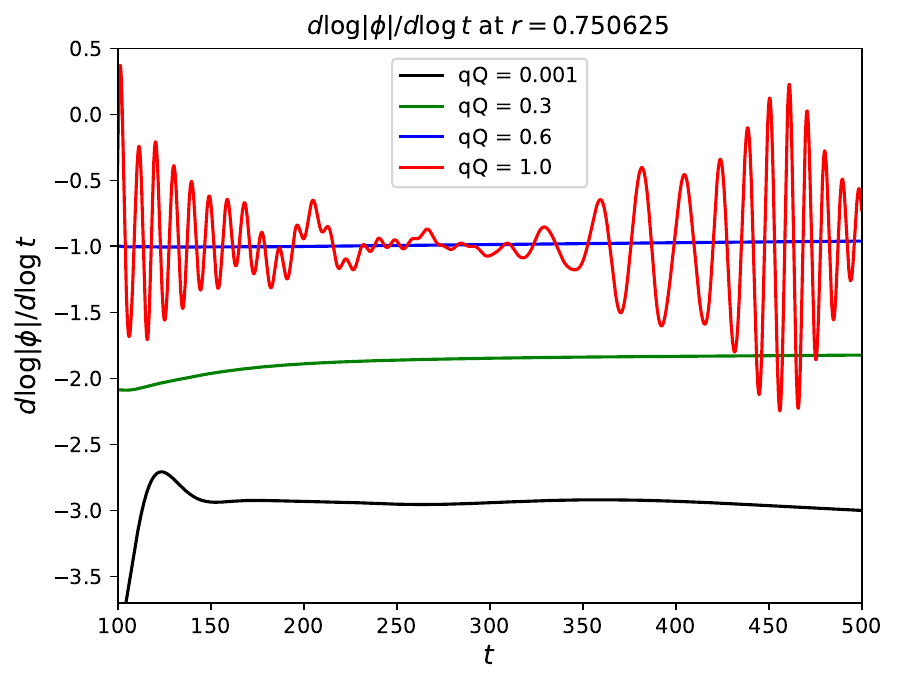}
    \caption{Derivative of the log of the absolute value of the scalar field with respect to the log of time. This gives the time exponent for each value of the charge parameter $Qq$, top for $\mathscr{I}^+$ and bottom for $r\approx 0.75$.}
    \label{fig:rnqnmlogp}
    \vspace{-3mm}
\end{figure}

To better analyze the tail decays' exponents \eqref{eq:chargedscalardecaytime}-\eqref{eq:chargedscalardecaynull}, we plot the following quantity,
\begin{equation}
    \frac{d \log |\phi|}{d\log t}\bigg|_{\text{late-time}} = \frac{d}{d\log t}\log \left( At^{p} \right) = p
    \label{eq:B}
\end{equation}
where $p$ is the exponent and $A$ is the amplitude of the scalar field at the point it enters the tail regime. The exponent $p$ attains in spherical symmetry, along a null hypersurface such as \scrip and at timelike infinity, the following expressions \cite{hod_late-time_1998_III}:
\begin{equation}
    p_{\scri^+} = \text{Re}(-\beta) - 1, \quad p_{i^+} = \text{Re}(-2\beta) {\high -} 2\,,
    \label{eq:severalp}
\end{equation}
where $\text{Re}(\beta)$ means the real part of $\beta$. \eqref{eq:chargedscalardecaytime} also tells us that there is a limit to the real part of $\beta$, when $4(qQ)^2 > 1$, i.e., when $qQ > 1/2$ or $qQ < -1/2$. It is important to note that our simulations do not reach $i^+$. They would if we were to leave the simulation to run for an infinite amount of time. However, it is expected that, as time increases, the decay of the scalar field at a given radius smaller than $r_{\scri^+}$ will tend to the decay rate at $i^+$. Therefore, from now on we compare the behavior of the scalar field at a given extraction radius of our choice with the decay rate $p_{i^+}$ given by the expression in \eqref{eq:severalp}. Plotting the data output from our simulation as in \eqref{eq:B}, we get Fig.~\ref{fig:rnqnmlogp}.

\begin{figure}[h]
    \centering
    \includegraphics[width=0.48\textwidth]{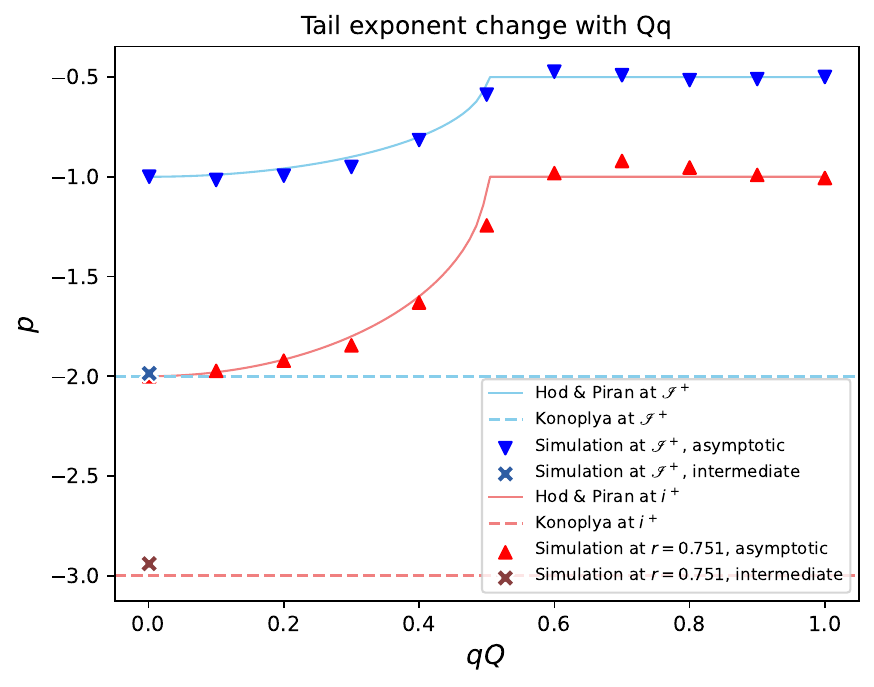}
        \caption{Comparison of the results from our simulation (the triangles {\high and crosses}) with the analytical results coming from Konoplya and Zhidenko \cite{konoplya_massive_2013} and from Hod and Piran \cite{hod_late-time_1998}. {\high Blue lines and symbols correspond to values at \scrip.} As mentioned in the main body of the text, the red lines correspond to the expected decay of the scalar field at $i^+$. The red triangles, however, correspond to the decay of the scalar field at a given radius $r \approx 0.75$ and at finite time, which is not $i^+$. Nonetheless, as the simulation time increases, the decay we expect to see at $r<1$ should approach asymptotically the one at $i^+$.}
    \label{fig:CompRNQNM}
    \vspace{-3mm}
\end{figure}

For $qQ = 0.6$ and $qQ=1.0$ both signals oscillate around $p = -0.5$ and $p=-1.0$ for \scrip and $r \approx 0.75$, respectively, meaning that the maximum real part of $\beta$ has been reached (check \eqref{eq:severalp} and \eqref{eq:chargedscalardecaytime}). To understand better how the exponent changes, we considered more values of $qQ$. To get precise values of $p$, we did a linear regression on $d\log|\phi|/d \log t${\high, whose slope provides the value of $p$. 
For $qQ\gtrsim 0.1$ performing the regression between 
$t=200$ and $t=500$ is valid, as the asymptotic regime is already reached at those times. However, in the limit $qQ\ll1$ the electromagnetic contribution dominates asymptotically for $t\gg\frac{M}{qQ}$ \cite{hod_late-time_1998,hod_late-time_1998_II}, so that simulations need to be evolved for longer to reach the desired regime. 
}
{\high In Fig.~\ref{fig:CompRNQNM} we show the results of several simulations, where we have fixed $M=q=1$, and we change the value of $Q$. This makes the overall factor $qQ$ change exactly in the same way as $Q$. The asymptotic exponents (i.e., respecting the condition $t\gg\frac{M}{qQ}$) extracted from our simulation are plotted in triangles. Solid lines represent the analytical predictions as shown in Ref.~\cite{hod_late-time_1998_III}, while dashed lines correspond to the predictions shown in Ref.~\cite{konoplya_massive_2013}. We also show the values of $p$ for the case when $Q = 0.001$ estimated between $t=200$ and $t=500$, represented through two crosses in the same color scheme. For this small value of $qQ$, these times correspond to the intermediate regime instead of the asymptotic one -- to obtain $p$ for the latter we ran the simulation until $t = 6000$.  It is interesting to note that initially, when the charged scalar field in the presence of a charged black hole enters the tail regime, it initially behaves as the neutral scalar field. This happens because the amplitude of the charged scalar field is mostly dominated initially by the real or complex part of the field, depending whether we perturb the former or the latter in the initial data. Only after some time, when both parts become comparable, does the asymptotic exponent corresponding to the charged scalar field appear. The main conclusion from this comparison is that our simulations coincide well with previous analytical results in the literature. Furthermore, in Fig.~\ref{fig:CompRNQNM}, we show not only the values of $p$ at \scrip (in blue), but also the values along a timelike hypersurface corresponding to the compactified value of $r = 0.751$ (in red). The main idea behind comparing the values of $p$ along these hypersurfaces to the values of $p$ as expected at $i_+$ is that as time progresses onwards, the observer moves ever closer to the position of $i_+$. It will never reach it, but it asymptotes to it. This is exactly what we see in the red quantities in Fig.~\ref{fig:CompRNQNM}.}

\begin{figure}[h]
    \centering
    \includegraphics[width=0.48\textwidth]{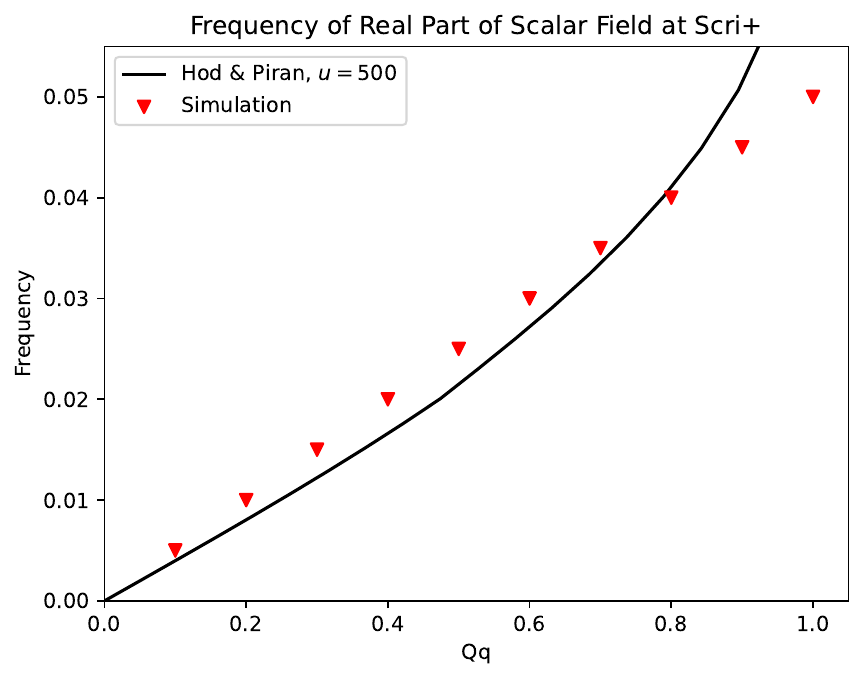}
        \caption{Frequency comparison of the results from our simulation (the triangles) with the analytical results coming from Hod and Piran \cite{hod_late-time_1998}, \eqref{eq:chargedscalardecaynull}, {\high where the black line corresponds to the frequency given by \eqref{eq:analytical_freq} divided by $2\pi$ and evaluated at $u=500$ (visually indistinguishable from the same expression evaluated at $u=2000$)}.} 
    \label{fig:CompFreqRNQNM}
    \vspace{-5mm}
\end{figure}

We also wanted to see the effect of $qQ$ in the oscillation frequency of the real and imaginary parts of the scalar field at \scrip. {\high The oscillatory part of~\eqref{eq:chargedscalardecaynull} is 
\begin{equation*}
\begin{aligned}
\phi_\textrm{oscil}=e^{i\varphi_\textrm{oscil}}&=e^{i\frac{qQ}{r_+}t}u^{i\left[qQ-\text{Im}(\beta)\right]} \\&= e^{i\frac{qQ}{2\,r_+}u}e^{i\left[qQ-\text{Im}(\beta)\right]\log(u)}\\&=e^{i\left\{\frac{qQ}{2\,r_+}u+\left[qQ-\text{Im}(\beta)\right]\log(u)\right\}} ,
\end{aligned}
\end{equation*}
so that its $u$-dependent angular frequency is 
\begin{equation}
\omega=\frac{d\varphi_\textrm{oscil}}{du} = \frac{qQ}{2\,r_+}+\left[(qQ-\text{Im}(\beta)\right]\frac{1}{u} . 
\label{eq:analytical_freq}
\end{equation}
}
To compare the analytical results with our simulation, we applied a Fourier transform to the real part (or equivalently to the imaginary part) of the scalar field during the tail regime. {\high This analysis is shown by the red triangles in Fig.~\ref{fig:CompFreqRNQNM}. The predicted frequency \eqref{eq:analytical_freq} (divided by $2\pi$ for ease of comparison) is evaluated at $u = 500$, i.e., the same value of the retarded time as that at which we are extracting the frequency. This is because our hyperboloidal time $t$ at \scrip and $u$ are comparable in the asymptotic regime. As $u$ increases, the shape of \eqref{eq:analytical_freq} becomes ever more dominated by the factor $qQ/(2r_+)$. We can therefore see that there is a good match between our results and the prediction by \eqref{eq:analytical_freq}, except for large values of $qQ$ close to extremality, where the divergence becomes quite stark and the data points from our simulation deviate. Understanding why this happens is left for future work.}

\section{Conclusions}
\label{sec:conclusions}
With this work, we took advantage of hyperboloidal slices to gain access to \scrip efficiently to understand the behavior of a charged scalar field perturbation of a RN black hole. This allowed us to study the QNM ringing of the scalar field in Schwarzschild and RN spacetimes and found it only possible to fit the spherical fundamental mode to the former case. The latter showed but one oscillation, making it difficult to construct a reliable fit. We propose that the reason for the almost complete absense of oscillations is due to the self interaction of the charged scalar field, which picks up as soon as the imaginary part is excited by the presence of the charged black hole. Regarding the tail regime, {\high our results agree very well with Ref.~\cite{hod_late-time_1998_III}}. 
Contrary to the point-of-view that electrically charged black holes might be astrophysically irrevelant \cite{shapiro_black_1983}, even if a black hole has a very small amount of charge, it may have a noticeable impact on the late-time behavior of perturbations.\par
We also analysed the frequency of oscillation of the real and imaginary parts during the tail regime and found {\high good agreement with the analytical predictions in Ref.~\cite{hod_late-time_1998_III}, although differences appear towards the extremal case.}

Natural extensions of this work will involve studying the decay tails that originate in a collapsing spacetime~\cite{Gundlach:1993tn} -- for a sample simulation of charged collapse with our infrastructure see Ref.~\cite{alvares_free_2025}. This will allow us to compare to prior results obtained at future null infinity~\cite{Purrer:2004nq}. The prospect of tuning the collapse towards the extremal case $Q=M$~\cite{Kehle:2024vyt} is especially exciting. 
Another option is to consider a scalar-field-dependent coupling with the Maxwell part such as Ref.~\cite{Guo:2025xwh} and study the behaviour of the resulting hairy black holes. Recent work~\cite{Cao:2024sot} used hyperboloidal slices to study quasinormal modes of Boulware-Deser-Wheeler black holes in Einstein-Gauss-Bonnet gravity. Given the potential ill-posedness of the latter, attempting a full non-linear simulation of this setup is too risky at this stage, but could be an interesting research line to follow in the future. 

\section*{Acknowledgements}
The authors would like to thank Christian Peterson Bórquez for useful advice regarding the ringing of QNM in hyperboloidal slices and to Adrien Kuntz and Nicola Franchini for the help finding important references for this work. The acknowledgements extend naturally to Vitor Cardoso and again to Adrien for very useful comments on the manuscript. We also thank the PRD referee for spotting a mistake in the equations that had gone undetected until now.
{\high Finally, we are indebted to Shahar Hod for 
pointing out several inaccuracies in an earlier version of
our paper.}\par
AVV thanks the Fundac\~ao para a  Ci\^encia e Tecnologia (FCT), Portugal, for the financial support to the Center for Astrophysics and Gravitation (CENTRA/IST/ULisboa) through the Grant Project~No.~UIDB/00099/2020. 
This work was also supported by the Universitat de les Illes Balears (UIB); the Spanish Agencia Estatal de Investigación grants PID2022-138626NB-I00, RED2022-134204-E, RED2022-134411-T, funded by MICIU/AEI/10.13039/501100011033 and the ERDF/EU; and the Comunitat Autònoma de les Illes Balears through the Conselleria d'Educació i Universitats with funds from the European Union - NextGenerationEU/PRTR-C17.I1 (SINCO2022/6719) and from the European Union - European Regional Development Fund (ERDF) (SINCO2022/18146). JDA thanks Fundação para a Ciência e Tecnologia (FCT), Portugal, for the financial support through the Grant Project  2024.04456.CERN.

\bibliographystyle{apsrev}
\bibliography{references}

\newpage

\end{document}